\documentclass[aps, prd, preprint, amsfonts, floatfix]{revtex4}

\usepackage{graphicx}
\usepackage{amsmath,amsfonts}
\usepackage{epsfig,color}

\begin{document}
\newcommand{\be}{\begin{eqnarray}}
\newcommand{\ee}{\end{eqnarray}}
\newcommand\del{\partial}
\newcommand\nn{\nonumber}
\newcommand{\Tr}{{\rm Tr}}
\newcommand{\Str}{{\rm Trg}}
\newcommand{\mat}{\left ( \begin{array}{cc}}
\newcommand{\emat}{\end{array} \right )}
\newcommand{\vect}{\left ( \begin{array}{c}}
\newcommand{\evect}{\end{array} \right )}
\newcommand{\tr}{{\rm Tr}}
\newcommand{\hm}{\hat m}
\newcommand{\ha}{\hat a}
\newcommand{\hz}{\hat z}
\newcommand{\hx}{\hat x}
\newcommand{\tm}{\tilde{m}}
\newcommand{\ta}{\tilde{a}}
\newcommand{\tz}{\tilde{z}}
\newcommand{\tx}{\tilde{x}}
\definecolor{red}{rgb}{1.00, 0.00, 0.00}
\newcommand{\rd}{\color{red}}
\definecolor{blue}{rgb}{0.00, 0.00, 1.00}
\definecolor{green}{rgb}{0.10, 1.00, .10}
\newcommand{\blu}{\color{blue}}
\newcommand{\green}{\color{green}}



\title{The Microscopic Twisted Mass Dirac Spectrum}
\author{K. Splittorff}
\affiliation{Discovery Center, The Niels Bohr Institute, University of Copenhagen, Blegdamsvej 17, DK-2100, Copenhagen {\O}, Denmark} 
\author{J.J.M. Verbaarschot}
\affiliation{Department of Physics and Astronomy, SUNY, Stony Brook,
 New York 11794, USA}

\date   {\today}
\begin  {abstract}

The microscopic spectral density for lattice QCD with two flavors and
maximally twisted mass is computed. The results are given for fixed
index of the Dirac operator and include the leading order $a^2$ 
corrections to the chiral Lagrangian due to the
discretization errors. The computation is carried out within the 
framework of Wilson chiral perturbation theory. 

\end{abstract}
\maketitle

\section{Introduction}

Large scale numerical simulations of twisted mass lattice QCD \cite{TM} 
are currently investigated in order to access the deep chiral regime of
QCD \cite{ETMC}. In the twisted mass formulation the standard Hermitian 
Wilson term is replaced by an anti-Hermitian isospin-violating Wilson 
term, for a review see \cite{Shindler}. 
Under an axial transformation, this modified Wilson term can be 
tranformed into the standard Wilson term while the mass is transformed 
into a twisted mass term. One advantage of this approach is that
the fermion determinant of the two flavor Dirac operator is bounded from
below due to the twisted quark mass. This offers automatic
control of the problems which the smallest eigenvalues of the Wilson
Dirac operator may cause for the numerical stability of standard
simulations with Wilson fermions.

The smallest eigenvalues of the Wilson Dirac operator also play a crucial role
for the spontaneous breakdown of chiral symmetry \cite{BC,Heller}. 
Here we compute
analytically the density of these smallest eigenvalues for twisted
mass lattice QCD. This microscopic eigenvalue density is uniquely
determined by the symmetries of the lattice theory and hence can be
obtained from the low energy effective theory known as Wilson chiral
perturbation theory \cite{SharpeSingleton,RS,BRS}. This effective theory describes the 
finite volume corrections as well as corrections due to discretization 
errors caused by the nonzero lattice spacing $a$. To 
order $a^2$, the effects of the lattice spacing are parametrized in terms 
of three additional low energy constants. While the value of these constants 
are specific to the exact implementation on the lattice, it is essential to 
know their values in order to extract physical observables such as the 
chiral condensate, $\Sigma$, and pion decay constant $F_\pi$. The analytical 
results for the eigenvalue density of the Dirac operator at nonzero twisted 
mass presented here offer a direct way to test Wilson chiral perturbation 
theory against lattice data. Moreover if the test is successful it 
provides a direct way to measure the additional low energy 
constants as well as the physical ones. Such a test has been carried out 
for  the quenched case with a standard (untwisted) mass in \cite{DWW,DHS}. 
Finally, we discuss constraints on the low energy constants from QCD inequalities.

This paper is organized as follows: To settle the notation the next section 
gives a brief introduction to twisted mass QCD. We then turn to the 
low energy effective theory, Wilson chiral perturbation theory, in section 
\ref{sec:WCPT}. The new results for the microscopic Dirac eigenvalue density 
at maximally twisted mass are presented in section \ref{sec:rho}. We discuss 
the constraints on the additional low energy constants from the perspective 
of QCD inequalites the in sections \ref{sec:QCDineq}.  Finally we draw conclusions in section \ref{sec:conc}.

\section{Basics of twisted mass QCD}
\label{sec:def}

Here we briefly recall the basics of twisted mass two flavor QCD in
the continuum limit as well as on the lattice, see \cite{TM} for
more details. This also introduces the notation used throughout this
paper.   

\subsection{Twisted mass in the continuum}

In the continuum formulation the twisted mass fermionic action is given by
\be
S = \int d^4x \ \bar{\psi}(D_\mu\gamma_\mu+m+iz_t\gamma_5\tau_3)\psi.
\ee
Under the axial transformation
\be
\psi'=\exp(i\omega\gamma_5\tau_3/2)\psi, \quad \quad  
\bar{\psi}'=\bar{\psi}\exp(i\omega\gamma_5\tau_3/2) 
\ee
the mass terms get rotated 
\be
\label{mass_rot}
m'&=&m\cos(\omega)- z_t\sin(\omega) \nn \\
z_t'&=&m\sin(\omega)+ z_t\cos(\omega), 
\label{mass-rot}\label{eq3}
\ee
as  follows from
\be
\exp(i\omega\gamma_5\tau_3) = \cos(\omega)+i\gamma_5\tau_3\sin(\omega).
\ee
The continuum covariant derivative term is of course invariant under
the axial transformation since $\{\gamma_5,\gamma_\mu\}=0$. In the continuum 
we therefore have
\be
\det(D_\mu\gamma_\mu+m+iz_t\gamma_5\tau_3) 
= \det(D_\mu\gamma_\mu+m'+iz_t'\gamma_5\tau_3). 
\ee    
Note that the twisted source, $z_t'$, vanishes completely if we make the
rotation with $\tan(\omega)=-z_t/m$.

If we simply want to evaluate the partition function at some
nonzero mass (and zero twisted mass) we could start with both 
$m$ and $z_t$ in the determinant as long as we remember that 
this corresponds to the value
\be
m'(\omega=\arctan(-z_t/m))=m\sqrt{1+\left(\frac{z_t}{m}\right)^2}  
\label{id}
\ee 
of the quark mass and zero value of the twisted mass.

Definition: {\sl Maximal twist} is obtained at $m=0$ with $\omega=\pi/2$. For maximal twist 
\be
\label{mp_zt}
m'&=& z_t \nn \\
z_t'&=& 0, 
\ee
so that
\be
Z(m=0,z_t;a=0) = Z(m'=z_t,z_t'=0;a=0). 
\label{eq8}
\ee

\subsection{Twisted mass Wilson fermions on the lattice}

With Wilson fermions on the lattice the discretized covariant derivative
\be
D_W =\frac{1}{2}\gamma_\mu(\nabla_\mu+\nabla_\mu^*)
     -\frac{ar}{2}\nabla_\mu\nabla_\mu^*
\ee
is {\sl not} anti-Hermitian and does {\sl not} anti-commute with $\gamma_5$. However, $D_W$ is $\gamma_5$-Hermitian 
\be
\gamma_5D_W\gamma_5=D_W^\dagger
\ee
and the product with $\gamma_5$, $D_5(m)\equiv\gamma_5(D_W+m)$ is therefore Hermitian. These properties are unaltered if one adds a clover term to $D_W$.

The main motivation to introduce the twisted mass becomes obvious when we
write the determinant in terms of the eigenvalues, $\lambda^5_j(m)$,
of $D_5(m)$ 
\be
\label{twisted_det}
\det(D_W+m+iz_t\gamma_5\tau_3)  
& = & \det(D_5(m)+iz_t\tau_3) \nn\\
&=&\prod_j(\lambda^5_j(m)+iz_t)(\lambda^5_j(m)-iz_t)=\prod_j(\lambda_j^{5}(m)^2+z_t^2).
\ee
The square of the twisted mass sets a lower limit on the terms in the
product even when the eigenvalues of $D_5(m)$  are smaller 
in magnitude than $m$ as happens for $a\neq0$. The numerical problem 
with small eigevalues of $D_5$ is therefore regulated by the twisted mass source.

Since the Wilson term breaks the axial-symmetry the identification 
(\ref{id}) for the partition function is no longer valid on the lattice. 
However, as the Wilson term is a cutoff artifact one is free to
choose the $m'$ as the physical quark mass provided that $z'=0$. Therefore,
if we start with $m=0$, it is natural to consider the twisted mass $z_t$ as
the phyiscal quark mass, cf. (\ref{mp_zt}). 

From Eq.~(\ref{twisted_det}) it is then clear that the Dirac spectrum 
relevant for chiral symmetry breaking at maximal twist is that of $D_5(m=0)$ 
\be
\frac{d}{dz_t} \log Z(m=0,iz_t,-iz_t;a) = \int d\lambda^5
\frac{2z_t }{\lambda^5(m=0)^2+z_t^2} \ \rho_5(\lambda^5(m=0),z_t;a). 
\ee
Note that, in the twisted-chiral limit we recover the Banks-Casher 
\cite{BC} relation \footnote{In practice, on the lattice it is challenging to reach these 
strict limits.} 
\be
\Sigma =\lim_{z_t\to0}\frac {\pi \rho_5(\lambda^5(m=0)=0;z_t;a)} V, 
\ee
It is therefore of particular interest to know the analytical form of 
$\rho_5(\lambda^5(m=0),z_t;a)$ in the microscopic limit. The microscopic 
eigenvalue density derived below gives exactly this form.


\section{Wilson Chiral Perturbation Theory with twisted mass}
\label{sec:WCPT}

With the twisted source included the static chiral Lagrangian reads 
\cite{SharpeSingleton,RS,BRS}
\be\label{Lstat}
V{\cal L}=\Tr(\hat{m}^\dagger U+\hat{m}U^\dagger) +\Tr(\hat{z}_t^\dagger
\tau_3 U-\hat{z}_t\tau_3U^\dagger) -\Tr(\hat{a}^\dagger
U\hat{a}^\dagger U+\hat{a} U^\dagger \hat{a} U^\dagger )
\ee
with the sources
\be
\hat{m}=m\Sigma V, \quad  \hat{z}_t=z_t\Sigma V 
\quad {\rm and} \quad  \hat{a}=aW_8V.
\ee
Here we have set $W_6=W_7=0$ \footnote{Note that we use the convention
  of \cite{DSV,ADSVprd} for the low energy constants $W_6$, $W_7$ and
  $W_8$. In \cite{BRS} these constants are denoted by $-{W_6}'$,
  $-{W_7}'$ and $-{W_8}'$ respectively.}. 

In the microscopic limit (aka. the $\epsilon$-regime) for twisted mass 
Wilson fermions \cite{Bar:2010zj} the zero momentum modes of
the pion fields factorize from the partition function resulting in the $z_t$-dependence
\be
Z_{2}^\nu(m,z_t;a)=\int_{U(2)}{\det}^\nu(U) \ e^{V{\cal L}}.
\ee   
Here we have written the expression for a sector with fixed index, $\nu$, 
of the Dirac operator. The index is defined through
\be
\nu\equiv\sum_{k} {\rm sign}\langle k|\gamma_5|k\rangle,
\ee 
where $|k\rangle$ are the eigenstates of $D_W$. Note that only the real modes of $D_W$ 
contribute to the index \cite{Itoh}. The index may also be obtained from the flow with 
$m$ of the eigenvalues of $D_5(m)$ \cite{HellerFlow}.

\section{The microscopic spectrum with two maximally twisted flavors}
\label{sec:rho}

Here we compute the microscopic spectral density of $D_5(m=0)$ relevant for 
two flavors at maximal twisted mass. The computation is carried out for fixed index, $\nu$, of the Wilson Dirac operator.

In order to derive this density we employ the graded generating functional with index $\nu$. This is given by \cite{DSV,ADSVprd,ADSVNf1}
 \be
\label{ZSUSY}
Z^\nu_{3|1}({\cal Z};a)  & = & \int \hspace{-1.5mm} dU \
{\rm Sdet}(iU)^\nu  
  e^{+\frac{i}{2}{\Str}({\cal Z}[U+U^{-1}])
    + a^2{{\Str}(U^2+U^{-2})}},  
\ee
where ${\cal Z}\equiv{\rm diag}(iz_t,-iz_t,z,\tilde{z})$, and the 
integration is over $Gl(3|1)/U(1)$. The difference with \cite{SV-Nf2} 
is that we now have the twisted mass instead of the standard mass. For a discussion
of the group manifold we refer to \cite{DOTV}.

The spectral resolvent is obtained from the graded generating functional by  differentiation with respect to 
the $z$ source and a subsequent quench of the additional flavors by the limit $z\to\tilde{z}$
\be
\label{Gsusy}
G^\nu_{3|1}(z,z_t;a)= \lim_{\tilde{z}\to z} \frac{d}{dz} {\cal Z}^\nu_{3|1}(iz_t,-iz_t,z,\tilde{z};a).
\ee
Finally, the density of eigenvalues, $\rho^\nu_5(\lambda^5,z_t;a)$, of $D_5$ follows from 
\be
\rho^\nu_5(\lambda^5,z_t;a) = \left \langle \sum_k \delta(\lambda^5_k-\lambda^5) \right \rangle_{N_f=2} = \frac{1}{\pi}{\rm Im}[G^\nu_{3|1}(z=-\lambda^5,z_t;a)]_{\epsilon\to0}.
\label{rho5def}
\ee

Our main task is therefore to evaluate the graded generating function. In \cite{SV-Nf2} it was shown that the generating functional
(\ref{ZSUSY}) can be rewritten as 
\be
Z_{3|1}^\nu({\cal Z}; a)
& =&  \frac{e^{-4  a^2}}{(16\pi a^2)^{2}}
\int_{-\infty}^\infty ds dt  \frac{B_{3|1}(S)}{B_{3|1}({\cal Z})}
 e^{(1/16 a^2){\rm Trg}(S^2+{\cal Z}^2)} e^{-t\tilde{z}/8 a^2}  
\det {e^{-is_k {\cal Z}_l/8 a^2}}_{k,l=1,2,3}
\nn \\
&& 
\times\int \hspace{-1.5mm} dU \ {\rm Sdet}(iU)^\nu 
e^{+ \frac i2 {\rm Trg}(SU + SU^{-1})} ,
\label{collectU}
\ee
where the Berezinian is given by
\be
B_{3|1}(S) = \frac {(is_3-is_2)(is_3-is_1)(is_2-is_1)}
{(t-is_1)(t-is_2)(t-is_3)}
\ee
and
\be
S\equiv \mat is &0 \\ 0 & t \emat
\label{sigma_diag}
\ee
with $s={\rm diag}(s_1,s_2,s_3)$.

The integral over $U$ results in the $a=0$ generating functional which
takes the form \cite{SplitVerb1,FA}
\be
\label{Z31a0}
Z^\nu_{3|1}(x_1,x_2,x_3,x_4;a=0) &=& 2 \frac{x_4^\nu}{x_1^\nu x_2^\nu x_3^\nu} 
\frac 1{(x_3^2-x_2^2)(x_3^2-x_1^2)(x_2^2-x_1^2)}
\\&& \hspace*{-6cm}\times
\det \left ( \begin{array}{cccc}
I_\nu(x_1)& x_1I_{\nu+1}(x_1) & x_1^2 I_{\nu+2}(x_1) & x_1^3 I_{\nu+3}(x_1) \\   
I_\nu(x_2)& x_2I_{\nu+1}(x_2) & x_2^2 I_{\nu+2}(x_2) & x_2^3 I_{\nu+3}(x_2)  \\ 
I_\nu(x_3)& x_3I_{\nu+1}(x_3) & x_3^2 I_{\nu+2}(x_3) & x_3^3 I_{\nu+3}(x_3)  \\ 
(-1)^\nu K_\nu(x_4)& x_4(-1)^{\nu+1}K_{\nu+1}(x_4) & x_4^2 (-1)^{\nu+2} K_{\nu+2}(x_4) & x_4^3 (-1)^{\nu+3} K_{\nu+3}(x_4)   
\end{array} \right ).\nn
\ee
We can thus write 
\be
Z_{3|1}^\nu({\cal Z}; a)
& =&  \frac{e^{-4  a^2}}{(16\pi a^2)^{2}}
\int ds dt \  \frac{B_{3|1}(S)}{B_{3|1}({\cal Z})} 
e^{(1/16 a^2){\rm Trg}(S^2+{\cal Z}^2)} e^{-t
 \tilde{z}/8 a^2}  \det( {e^{-is_k {\cal Z}_l/8 a^2}})_{k,l=1,2,3}\nn\\
&& 
\times \left ( \frac{\prod_k (-is_k)}{-t} \right )^{\nu}
 Z_{3|1}^\nu\left (\{(s_k^2)^{1/2}\} ,(-t^2)^{1/2}
; a= 0 \right ).
\label{collect}
\ee
The next step is to simplify the determinant
\be
\det(e^{-is_k{\cal Z}_j/8\hat{a}^2})_{k,j=1,2,3}=\left|\begin{array}{ccc}
e^{-is_1{\cal Z}_1/8\hat{a}^2} & e^{-is_1{\cal Z}_2/8\hat{a}^2} & e^{-is_1{\cal Z}_3/8\hat{a}^2}\\
e^{-is_2{\cal Z}_1/8\hat{a}^2} & e^{-is_2{\cal Z}_2/8\hat{a}^2} & e^{-is_2{\cal Z}_3/8\hat{a}^2}\\
e^{-is_3{\cal Z}_1/8\hat{a}^2} & e^{-is_3{\cal Z}_2/8\hat{a}^2} & e^{-is_3{\cal Z}_3/8\hat{a}^2}
\end{array}\right|.
\ee 
Since the other terms in the integrand also combine into an anti-symmetric function of the $s_k$, all terms in
the expansions of the determinant as a sum over permutations give the same contributions. In the integrand, we can thus make the replacement  
\be
\det(e^{-is_k{\cal Z}_j/8\hat{a}^2})_{k,j=1,2,3} \to
6e^{-is_1{\cal Z}_1/8\hat{a}^2-is_2{\cal Z}_2/8\hat{a}^2-is_3{\cal Z}_3/8\hat{a}^2}.
\ee 
The factor $e^{-i(s_1{\cal Z}_1+s_2{\cal Z}_2+s_3{\cal Z}_3)/8\hat{a}^2}$ is 
absorbed into the mixed term in the exponent of 
$e^{-\frac 1{16a^2}{\rm Trg}(S-{\cal Z})^2}$.
The inverse Berezinian of  ${\cal Z}$  becomes
\be
\frac{1}{B_{3|1}({\cal Z})}=\frac{(\tilde{z}-z_1)(\tilde{z}-z_2)(\tilde{z}-z)}{(z-z_1)(z-z_2)(z_2-z_1)}=\frac{(\tilde{z}-iz_t)(\tilde{z}+iz_t)(\tilde{z}-z)}{(z-iz_t)(z+iz_t)(-2iz_t)}.
\ee
This contributes a  total factor of $i/2z_t$ to the resolvent
$G$ (ie. after differentiation with respect to  $z$, and the
limit $\tilde{z}\to z$ has been taken, so that we necessarily have to differentiate the factor $(\tilde z - z)$).

Combining the above expressions the resolvent for $D_5(m=0)$ takes the form
\be\label{G31v2}
&& G^\nu_{3|1}(z,m=0,z_t;a) \\ &=& \frac i{\pi^2(16a^2)^2 z_t Z^\nu_{N_f =2}(iz_t,-iz_t;a)}
\int ds_1 ds_2 ds_3 dt \ 
\nn\\ &&\times
\frac{(is_2-is_1)(is_3-is_1)(is_3-is_2)} 
{(t-is_1)(t-is_2)(t-is_3)}\nn \\ &&\times
e^{-\frac 1{16a^2}[(s_1-z_t)^2+(s_2+z_t)^2 +(s_3+iz)^2 +(t-z)^2]  }
\frac{(is_1)^\nu(is_2)^\nu(is_3)^\nu}{(t)^\nu}
\nn\\&&\times
Z^\nu_{3|1}((s_1^2)^{1/2},(s_2^2)^{1/2},(s_3^2)^{1/2},(-t^2 )^{1/2};a=0),\nn
\ee
where the partition function for $a=0$ is given by Eq.~(\ref{Z31a0}) and the integration of $s_3+iz$ is over the real axis. The microscopic 
eigenvalue density of $D_5$ in the theory with two flavors at maximally twisted mass 
follows from (\ref{rho5def}). We only need to evaluate the two flavor maximally twisted mass 
partition function which appears in the normalization of $G_{3|1}$.

\subsection{The microscopic two flavor maximally twisted mass partition function}

In order to complete the computation of the microscopic eigenvalue density we need to evaluate the normalization which is given by the two flavor maximally twisted mass partition function
\be
\label{Z2}
Z^\nu_{2}(iz_t,-iz_t;a)  & = & \int_{U(2)} \hspace{-1.5mm} dU \
{\rm det}(iU)^\nu  
  e^{+\frac{i}{2}{\tr}({\cal Z}[U+U^{-1}])
    + a^2{{\tr}(U^2+U^{-2})}},  
\ee
where ${\cal Z}\equiv{\rm diag}(iz_t,-iz_t)$. Extending the results of  \cite{SV-Nf2} to the twisted mass case we find
\be
Z^\nu_{2}(iz_t,-iz_t;a) &= &\frac{i e^{4a^2}}{z_t \pi (16a^2)}
\int_{-\infty}^\infty\int_{-\infty}^\infty ds_1 ds_2 
(is_1-is_2)
  \\
&&\hspace{-2cm}\times e^{-\frac 1{4a^2}[(s_1-z_t)^2+ (s_2+z_t)^2]} (is_1)^\nu(is_2)^\nu 
Z^\nu_2(s_1,s_2;a=0), \nn
\label{ztwo_twist}
\ee
where
\be
\label{Z2a0}
Z^\nu_2(x_1,x_2;a=0) = \frac 2{x_1^\nu x_2^\nu (x_2^2-x_1^2)} \det \left | 
\begin{array}{cc} I_\nu(x_1) & x_1 I_{\nu+1}(x_1) 
\\ I_\nu(x_2)& x_2 I_{\nu+1}(x_2)
\end{array}  \right | .
\ee

The final step in the calculation is to factorize the four dimensional integrals 
in Eq. (\ref{G31v2}) into the product of two-dimensional integrals.
Not only may this factorized form have a deep connection to an underlying 
integrable hierarchy \cite{Toda}, it is also highly advantageous for numerical 
evaluation of the eigenvalue density. 

\vspace{3mm}

In \ref{app:factor} we show that the spectral resolvent (\ref{G31v2}) for 
the microscopic eigenvalue density of $D_5$ with two flavors at maximally 
twisted mass can be written as
\be
\label{G31factorized}
G_{3|1}^\nu(z,z_t;a) & = & 
G_{1|1}^\nu(z,z;a)
\\
&& +\frac{Z_2(iz_t,z;a)}{Z_2^\nu(iz_t,-iz_t;a)}
\frac{z-iz_t}{2iz_t} G_{1|1}^\nu(-iz_t,z;a) 
-\frac{Z_2^\nu(-iz_t,z;a)}{Z_2^\nu(iz_t,-iz_t;a)} 
\frac{z+iz_t}{2iz_t} G_{1|1}^\nu(iz_t,z;a). \nn
\ee
Here 
\be
G_{1|1}^\nu(z_1,z_2;a) & = & -\frac{1}{16 a^2 \pi} \int_{-\infty}^\infty dsdt \
   \frac{1}{t+z_2-is-z_1} e^{-(s^2 + t^2)/(16 a^2)} 
\nn \\
&& \hspace{1cm}
\times \left(\frac{is+z_1}{t + z_2}\right)^\nu 
Z_{1|1}^\nu(\sqrt{-(is + z_1)^2},\sqrt{-(t + z_2)^2},a=0)
\label{G11gen}
\ee
with
\be
Z_{1|1}^\nu(m_1,m_2;a=0) = \left(\frac{m_2}{m_1}\right)^\nu (I_\nu(m_1) m_2 K_{\nu+1}(m_2) 
     +m_1 I_{\nu+1}(m_1)K_{\nu}(m_2))
\ee
and
\be
Z_2^\nu(z_1,z_2;a) & = & 
 \frac{1}{\pi 16 a^2} \int_{-\infty}^\infty ds_1 ds_2 \  
    \frac{1}{(z_2 - z_1)} (is_1+z_1-is_2-z_2) e^{-(s_1^2+s_2^2)/(16 a^2)}
\nn \\
&& \hspace{1cm}
\times \left(\frac{is_1+z_1}{is_2+z_2}\right)^\nu 
Z_2^\nu(\sqrt{-(is_1+z_1)^2},\sqrt{-(is_2+z_2)^2};a=0)
\label{Z2gen}
\ee
with $Z_2^\nu(x_1,x_2;a=0)$ given in Eq. (\ref{Z2a0}).
Note that the first term on the right hand side of (\ref{G31factorized}) gives rise to the quenched density of $D_5$ at zero untwisted mass, $m$. A similar factorization of the unquenched density has been observed in the microscopic limit of QCD at nonzero chemical potential \cite{AOSV}. In that case this structure has been understood in terms of an underlying integrable hierarchy.

With Eq.~(\ref{G31factorized}) the 
spectral density has been expressed in terms of products of double integrals. 
This form is far easier to evaluate numerically than the four fold integral 
given in Eq.~(\ref{G31v2}).

This completes the computation of the microscopic eigenvalue density of $D_5(m=0)$ for 
two flavors at maximal twisted mass in sectors with fixed index of the Wilson Dirac 
operator. See Figure \ref{fig:twisted} for plots of the density. Note in particular the 
behavior of the near zero-modes.   

\begin{center}\vspace{1cm}
\begin{figure}[t*]
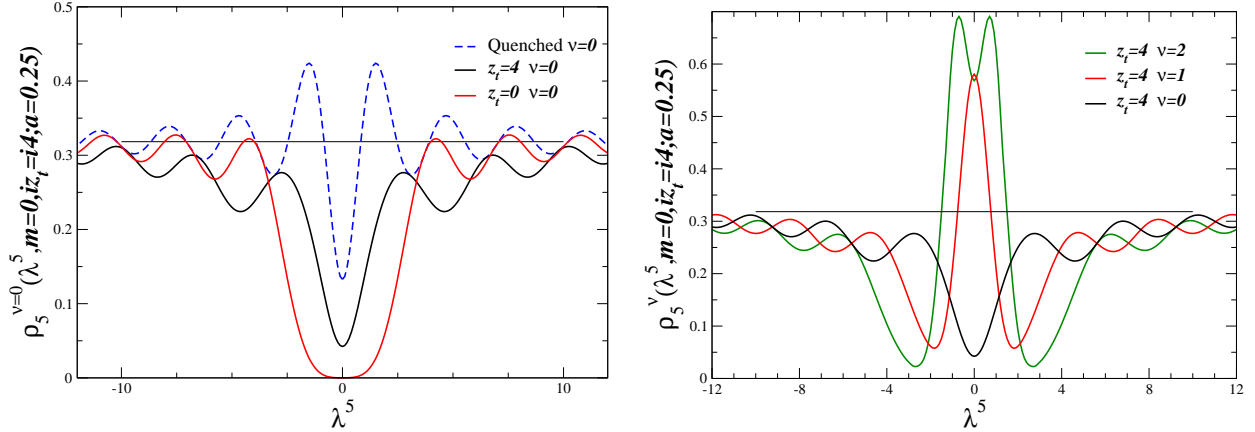

\includegraphics[width=8cm,angle=0]{Wilson-twisted-mt4a0p25.eps}
\hfill
\includegraphics[width=8cm,angle=0]{Wilson-twisted-mt4a0p25-nu1.eps}
\caption{\label{fig:twisted} The spectrum of $D_5(m=0)$ for two
  flavors with maximally twisted mass. {\bf Left:} the sector with 
  zero index of the Dirac operator. As the twisted mass increases the 
  quenched result (dashed curve) is approached. {\bf Right:} the dependence 
  on the index $\nu$ for fixed $z_t=4$ and $a=0.25$. As $W_8$ is decreased 
  the near zero-modes become exact $\delta$-functions at $\lambda^5=0$. For 
  small lattice spacing the width of the peak is proportional to $\sqrt{W_8}$.
  The thin horizontal line in both plots indicates the value $1/\pi$ which 
  is the asymptotic limit of the density for large values of $|\lambda_5|$.}
\end{figure}
\end{center}

\section{QCD inequalities with twisted quark mass}
\label{sec:QCDineq}

In this section we discuss two QCD inequalities. First, a QCD inequality for 
the microscopic partition function in a sector with fixed $\nu$ and second 
a QCD inequality for the pion masses. We will see that both put constraints 
on the low energy constants of Wilson chiral perturbation theory.

The twisted mass $N_f=2$ QCD partition function is positive 
definite for all $\nu$. This imposes a positivity requirement
of the  partition function corresponding the chiral Lagrangian
of the Wilson QCD partition function. Because of the identity
\be
Z_{2}^\nu(z_t=0;W_6,W_7,W_8,a) = (-1)^{\nu}Z_{2}^\nu(z_t=0;-W_6,-W_7,-W_8,a),
\ee
and because for large $z_t$, the sign of the partition function is independent
of the $W_k$, we necessarily obtain contraints on the $W_k$. In case
$W_6=W_7 =0$ we find that $W_8> 0$. From the small $a$-expansion of the 
partition function we obtain the condition
\be
W_8-W_6-W_7 > 0,
\ee
in agreement with the convergence requirements of the graded partition function
\cite{ADSVprd}. Additional constraints can be obtained from mass inequalities for the pion masses
which will be discussed in the remainder of this section.

The Dirac operator including the twisted mass,
\be
D_W + m + i z \tau_3 \gamma_5 ,
\ee
has the Hermiticity property 
\be
\tau_1 \gamma_5 (D_W + m + i z \tau_3 \gamma_5) \gamma_5 \tau_1 = (D_W + m + i z \tau_3 \gamma_5)^\dagger. 
\label{twisted-H-rel}
\ee
Therefore the inverse Dirac operator 
\be
    S(x,y)=\langle x |\frac 1 {D_W + m + i z \tau_3 \gamma_5}|y\rangle
\ee
satisfies 
\be
S(x,y)^\dagger =\gamma_5 \tau_1 S(y,x) \tau_1 \gamma_5.
\ee
Instead of $\tau_1$  we could of course also have used 
\be
\cos(\phi)\tau_1+\sin(\phi)\tau_2
\ee
in the Hermiticity relation (\ref{twisted-H-rel})
which leads to the same consequences. All we need is a
 combination that anticommutes with $\tau_3$ and is unitary.
This relation allows us to derive Weingarten type inequalities
\cite{Weingarten,Witten,SS} for the pion masses.

The correlation function of two messon sources
$\bar{\psi}\Gamma \psi(x)$
and $\bar{\psi}\Gamma \psi(y)$ evaluated for
a fixed background gauge field satisfies ($\Gamma$ is unitary)
\be
 \label{QCDineqFixedG}        
\int {\cal D} \, \bar\psi  {\cal D} \, \psi \ \bar\psi(x)\Gamma \psi(x) \bar \psi(y) \Gamma \psi(y)  
      & = & - \Tr[ S(y,x) \Gamma S(x,y)  \Gamma] +\Tr[ S(x,x) \Gamma]\Tr[ S(y,y) \Gamma]
                                \\
        & = & \Tr[S(y,x) \Gamma i\tau_1\gamma_5 S(y,x)^\dagger
            i\tau_1\gamma_5 \Gamma
                              ]+\Tr[ S(x,x) \Gamma]\Tr[ S(y,y) \Gamma]
                                \nn \\
             & \leq &  \Tr[ S(y,x) S(y,x)^\dagger] +\Tr[ S(x,x) \Gamma]\Tr[ S(y,y) \Gamma]
\ .\nn
                                 \ee
            The bound in the inequality is saturated for
$\Gamma=i\gamma_5\tau_1$ (or with $\tau_1\to\tau_2$ but {\sl not} with
$\tau_1\to\tau_3$). 
This inequality 
has been  evaluated for a fixed
  gauge field background. However, since the fermion determinant is
           positive for all gauge field configurations the
             inequality continues to hold after averaging. 
If the disconnected diagrams average to zero, we obtain
                                 \be
\langle 
\int {\cal D} \, \bar\psi  {\cal D} \, \psi \ 
 \bar\psi(x)\Gamma \psi(x) \bar \psi(y) \Gamma \psi(y)        
   \rangle 
& \leq &  \langle \Tr \, S(0,x) S(0,x)^\dagger \rangle \ .
\label{twistedQCDineq} 
\ee 
For  mesonic channels with  mass gap $m_\Gamma$ we have 
\be
\langle \int {\cal D} \, \bar\psi  {\cal D} \, \psi \ 
\bar\psi(x)\Gamma \psi(x) \bar \psi(0) \Gamma \psi(0)
\rangle \propto \exp(- m_\Gamma |x|) \ \ \ {\rm as} \ \ \ x \to \infty   \ .
\ee
The inequality for the correlators thus translates into an inequality 
for the meson masses. From
(\ref{twistedQCDineq}) we then conclude that \cite{talk-ect}
\be
m_{i\gamma_5\tau_{1,2}} \leq m_\Gamma \ .
\label{Mineq2}
\ee
In particular, we have 
\be
m_{\pi^\pm} \leq m_{\pi^0}.
\ee
From leading order Wilson chiral perturbation theory one obtains
\cite{hep-lat/0407025,hep-lat/0407006}
\be
 (m_\pi^0)^2 -(m_\pi^\pm)^2 = \frac {16 a^2(W_8+2W_6)}{F_\pi^2}.
\label{pions}
\ee 
If the contribution from disconnected diagrams is not important we conclude
that 
\be
W_8+2W_6 > 0.
\ee
The contribution of the disconnected diagrams can be isolated by the introduction of valence quarks. This results in the inequality \cite{arXiv:1111.2404}
\be
W_8 > 0,
\label{sharpe}
\ee
independent of the value of $W_6$ and $W_7$. Lattice simulations for 
twisted mass fermions in \cite{arXiv:0911.5061} show that 
\be
m_\pi^0 < m_\pi^\pm.
\ee
This implies that the contribution of the disconnected diagrams is important
for the simulations in \cite{arXiv:0911.5061}. The possible importance of disconnected diagrams has been studied explicitly in lattice simulations of the
respective correlators in \cite{arXiv:0709.4564}. Using Eq. (\ref{pions}) we thus conclude that for the simulations in \cite{arXiv:0911.5061}
\be
W_8+2 W_6 < 0.
\ee
Combined with the inequality (\ref{sharpe}) derived in \cite{arXiv:1111.2404} we
obtain the constraint
\be
W_6 <0.
\ee

In the quenched case lattice simulations show that the charged
pions are the lightest pseudoscalar Goldstone bosons \cite{757044}. This is an agreement with the lore that disconnected diagrams are suppressed in the quenched theory \cite{Witten}.

\section{Conclusions}
\label{sec:conc}

We have computed the microscopic spectral density of the massless Hermitian Wilson Dirac 
operator in the presence of two dynamical flavors at nonzero maximally twisted mass. 
The characteristic shape of the eigenvalue density in sectors with fixed index of 
the Wilson Dirac operator derived in this paper offers a direct way to test Wilson chiral 
perturbation theory for twisted mass against lattice QCD. If the spectral density obtained on 
the lattice follows the analytical prediction, the strong dependence of the 
analytical result on the low energy constant $W_8$ offers a direct way to 
measure the value of $W_8$. We have reduced the analytical form of the
twisted mass microscopic spectral density to a factorized form that is
easily evaluated with standard numerical methods. A similar factorized
form of the density for two standard dynamical flavors was recently
presented in \cite{SV-lat2011}.

The microscopic results for the spectral density of $D_5$ at $m=0$ have been derived for $W_8>0$. 
As has been argued in \cite{DSV,ADSVprd} only the theory with $W_8>0$ correctly describes 
lattice QCD with Wilson fermions. In support of this we have checked that the microscopic 
partition function for two flavors at maximally twisted mass is a positive definite function 
in all sectors with fixed index $\nu$ of the Wilson Dirac operator.

\noindent
{\bf Acknowledgments:}
We would like to thank Roberto Frezzotti, Giancarlo Rossi, as well as
the participants of the ECT$^*$ workshop 'Chiral dynamics with Wilson
fermions' for useful discussions. In particular, we would like to acknowledge 
discussions with Gregorio Herdoiza at UAM, may 2011, on inequlites for 
twisted mass QCD. We would like to thank the Laboratori Nazionali di Frascati 
for hospitality during the early stages of this project. This work was supported by U.S. 
DOE Grant No. DE-FG-88ER40388 (JV) and the {\sl Sapere Aude} program of The Danish
Council for Independent Research (KS).


\renewcommand{\thesection}{Appendix \Alph{section}}
\setcounter{section}{0}

\section{factorization of $G_{3|1}$}
\label{app:factor}

In this Appendix we show that the microscopic eigenvalue density for two flavors of 
maximally twisted mass can be factorized into two-dimensional integrals. 

We start from the resolvent which is given by Eq.~(\ref{G31v2})
\be
&& G_{3|1}(z,m=0,z_t; a) \nn\\
&=& \frac 1{\pi^2(16a^2)^2 Z^\nu_{N_f=2}(iz_t,-iz_t;a)} \int ds_1ds_2ds_3dt \frac i{z_t} \frac {(is_2-is_1)(is_3-is_1)(is_3-is_2)}{(t-is_1)(t-is_2)(t-is_3)}\\
&&\times e^{-[(s_1-z_t)^2 +(s_2+z_t)^2+(s_3+iz)^2+(t-z)^2]/16a^2}
\frac{(is_1 is_2 is_3)^\nu}{t^\nu} \nn\\
&&\times Z^\nu_{3|1}((s_1^2)^{1/2},(s_2^2)^{1/2},(s_1^2)^{1/2},(-t^2)^{1/2};a=0),\nn
\ee
and use the notation
\be
x_k = (s_k^2)^{1/2}, \quad k =1, 2, 3, \qquad x_4 = it.
\ee
Our aim is to rewrite this in a factorized form. To this end we explicitly insert the 
$a=0$ partition function given in (\ref{Z31a0}) and consider the combination
\be
&& \frac {(is_2-is_1)(is_3-is_1)(is_3-is_2)}{(t-is_1)(t-is_2)(t-is_3)}
\frac 1{(x_3^2-x_2^2)(x_3^2-x_1^2)(x_2^2-x_1^2)} \\
&&\times\det \left ( \begin{array}{cccc}
I_\nu(x_1)& x_1I_{\nu+1}(x_1) & x_1^2 I_{\nu+2}(x_1) & x_1^3 I_{\nu+3}(x_1) \\   
I_\nu(x_2)& x_2I_{\nu+1}(x_2) & x_2^2 I_{\nu+2}(x_2) & x_2^3 I_{\nu+3}(x_2)  \\
  I_\nu(x_3)& x_3 I_{\nu+1}(x_3) & x_3^2  I_{\nu+2}(x_3) & x_3^3  I_{\nu+3}(x_3)
  \\ 
(-1)^\nu K_\nu(x_4)& x_4(-1)^{\nu+1}K_{\nu+1}(x_4) & x_4^2
  (-1)^{\nu+2} K_{\nu+2}(x_4) & x_4^3 (-1)^{\nu+3} K_{\nu+3}(x_4)
\end{array} \right) \vspace*{1cm} . \nn
\ee
Combining the prefactors and using recursion relations for Bessel functions
this can be rewritten as
\be
&&\frac 1{(x_1+x_2)(x_1+x_3)(x_1+x_4)(x_2+x_3)(x_2+x_4)(x_3+x_4)} \\
&&\times\det \left ( \begin{array}{cccc}
I_\nu(x_1)& x_1I_{\nu+1}(x_1) & x_1^2 I_{\nu}(x_1) & x_1^3 I_{\nu+1}(x_1) \\   
I_\nu(x_2)& x_2I_{\nu+1}(x_2) & x_2^2 I_{\nu}(x_2) & x_2^3 I_{\nu+1}(x_2)  \\
  I_\nu(x_3)& x_3 I_{\nu+1}(x_3) & x_3^2  I_{\nu}(x_3) & x_3^3  I_{\nu+1}(x_3)
  \\ 
(-1)^\nu K_\nu(x_4)& x_4(-1)^{\nu+1}K_{\nu+1}(x_4) & x_4^2 (-1)^{\nu}
  K_{\nu}(x_4) & x_4^3 (-1)^{\nu+3} K_{\nu+1}(x_4)  \end{array}
\right) . \nn
\ee 
The factorized form is due to the appearance of $I_\nu / K_\nu$ in the odd 
columns and $I_{\nu+1} / K_{\nu+1}$ in the even colums. 
Expanding the determinant results in
\be
&&\frac 1{(x_1+x_2)(x_1+x_3)(x_1+x_4)(x_2+x_3)(x_2+x_4)(x_3+x_4)}\nn \\
&\times&\left [
 -(-1)^{\nu+1}I_{\nu+1}(x_3)K_{\nu+1}(x_4)I_{\nu}(x_1)I_{\nu}(x_2)x_3x_4(x_3^2-x_4^2)(x_1^2-x_2^2) \right .\nn\\&&
 +(-1)^\nu K_{\nu+1}(x_4)I_{\nu+1}(x_1)I_{\nu}(x_3)I_{\nu}(x_2)x_1x_4(x_1^2-x_4^2)(x_2^2-x_3^2) \nn\\&&
 -(-1)^{\nu}K_{\nu+1}(x_4)I_{\nu+1}(x_2)I_{\nu}(x_3)I_{\nu}(x_1)x_2x_4(x_2^2-x_4^2)(x_1^2-x_3^2)\nn\\&&
 +(-1)^{\nu+1}I_{\nu+1}(x_3)I_{\nu+1}(x_2)K_{\nu}(x_4)I_{\nu}(x_1)x_2x_3(x_2^2-x_3^2)(x_1^2-x_4^2)\nn\\&&
 -(-1)^{\nu+1}I_{\nu+1}(x_3)I_{\nu+1}(x_1)K_{\nu}(x_4)I_{\nu}(x_2)x_1x_3(x_1^2-x_3^2)(x_2^2-x_4^2)\nn \\&&
\left .  -(-1)^\nu I_{\nu+1}(x_1)I_{\nu+1}(x_2)K_{\nu}(x_4)I_{\nu}(x_3)x_1x_2(x_1^2-x_2^2)(x_3^2-x_4^2)\right ]. 
\ee
We then decompose the fractions as
\be
&& \frac {(x_1^2-x_4^2)(x_2^2-x_3^2)}{(x_1+x_2)(x_1+x_3)(x_1+x_4)(x_2+x_3)(x_2+x_4)(x_3+x_4)}\\
 & =& \frac 1{(x_1+x_2)(x_3+x_4)}-\frac 1{(x_1+x_3)(x_2+x_4)} \nn
\ee
and any cyclic permuatations thereof. This results in
\be
&& \frac 1{(x_1+x_3)(x_2+x_4)} \nn\\
&\times&[-(-1)^{\nu+1}I_{\nu+1}(x_3)K_{\nu+1}(x_4)I_{\nu}(x_1)I_{\nu}(x_2)x_3x_4
-(-1)^{\nu}K_{\nu+1}(x_4)I_{\nu+1}(x_1)I_{\nu}(x_3)I_{\nu}(x_2)x_1x_4\nn\\
&&-(-1)^{\nu+1}I_{\nu+1}(x_3)I_{\nu+1}(x_2)K_{\nu}(x_4)I_{\nu}(x_1)x_2x_3
-(-1)^\nu I_{\nu+1}(x_1)I_{\nu+1}(x_2)K_{\nu}(x_4)I_{\nu}(x_3)x_1x_2 ]\nn\\ 
&&+\frac 1{(x_2+x_3)(x_1+x_4)} \nn \\
&\times&[(-1)^{\nu+1}I_{\nu+1}(x_3)K_{\nu+1}(x_4)I_{\nu}(x_1)I_{\nu}(x_3)x_4x_4 
+(-1)^\nu K_{\nu+1}(x_4)I_{\nu+1}(x_2)I_{\nu}(x_3)I_{\nu}(x_1)x_2x_4]\nn \\&&
+(-1)^{\nu+1}I_{\nu+1}(x_3)I_{\nu+1}(x_1)K_{\nu}(x_4)I_{\nu}(x_2)x_1x_3
+(-1)^\nu I_{\nu+1}(x_1)I_{\nu+1}(x_2)K_{\nu}(x_4)I_{\nu}(x_3)x_1x_2  ]\nn\\
&&+\frac 1{(x_1+x_2)(x_3+x_4)} \nn\\
&\times&[(-1)^\nu K_{\nu+1}(x_4)I_{\nu+1}(x_1)I_{\nu}(x_3)I_{\nu}(x_2)x_1x_4
-(-1)^{\nu}K_{\nu+1}(x_4)I_{\nu+1}(x_2)I_{\nu}(x_3)I_{\nu}(x_1)x_2x_4
\nn \\ &&  (-1)^{\nu+1}I_{\nu+1}(x_3)I_{\nu+1}(x_2)K_{\nu}(x_4)I_{\nu}(x_1)x_2x_3 
-(-1)^{\nu+1}I_{\nu+1}(x_3)I_{\nu+1}(x_1)K_{\nu}(x_4)I_{\nu}(x_2)x_1x_3 ].\nn 
\ee
\be
= 
\frac{(-1)^\nu(x_4 K_{\nu+1}(x_4)I_\nu(x_2)+x_2K_\nu(x_4)I_{\nu+1}(x_2)}{x_4+x_2}\frac{x_3I_{\nu+1}(x_3)I_\nu(x_1) -x_1I_{\nu+1}(x_1)+I_\nu(x_3) }{x_1+x_3}\nn\\
\frac{(-1)^\nu(x_4 K_{\nu+1}(x_4)I_\nu(x_1)+x_1K_\nu(x_4)I_{\nu+1}(x_1)}{x_4+x_1}\frac{x_2I_{\nu+1}(x_2)I_\nu(x_3) -x_3I_{\nu+1}(x_2)I_\nu(x_2) }{x_2+x_3}\nn\\
\frac{(-1)^\nu(x_4 K_{\nu+1}(x_4)I_\nu(x_3)+x_3K_\nu(x_4)I_{\nu+1}(x_3)}{x_3+x_4}\frac{x_1I_{\nu+1}(x_1)I_\nu(x_2) -x_2I_{\nu+1}(x_2)+I_\nu(x_1) }{x_1+x_2}.\nn\\
\ee
Using this identity we can express the resolvent in the factorized form
\be
&& G^\nu_{3|1}(z,m=0,iz_t,-iz_t;a) \\
&=& G_{1|1}^\nu(z,z;a)
+\frac{Z_2^\nu(iz_t,z)(z-iz_t)}{Z_2^\nu(iz_t,-iz_t)2iz_t}G_{1|1}^\nu(-iz_t,z;a)  
-\frac{Z_2(-iz_t,z) (z+iz_t)}{Z_2(iz_t,-iz_t)2iz_t}G_{1|1}^\nu(iz_t,z;a), \nn
\ee
where $Z_2^\nu(z_1,z_2)$ and $G_{1|1}^\nu(z_1,z_2;a)$ are given in (\ref{Z2gen}) and 
(\ref{G11gen}) respectively. This factorization can also be derived in 
general terms \cite{Mario,KSV}.



\end{document}